\documentclass[10pt]{article}
\usepackage[margin=1in]{geometry}
\usepackage{amsmath, amssymb, amsfonts}
\usepackage{graphicx}
\usepackage{authblk}
\usepackage{physics}
\usepackage{hyperref}
\usepackage{bm}
\usepackage{xcolor}
\usepackage{amsmath}
\usepackage{dsfont,bbold}
\usepackage{mathtools}
\usepackage{tikz}
\usetikzlibrary{arrows.meta, positioning, fit}
\usepackage{subcaption}
\usepackage[numbers]{natbib}
\usepackage{upgreek}
\usepackage{comment}

\usepackage[scr=boondox,  
            frak = euler]   
           {mathalpha}

\title{A Programmable Linear Optical Quantum Reservoir with Measurement Feedback for Time Series Analysis}

\author[]{Çağın Ekici}

\date{\today}

\begin{document}

\maketitle
\begin{abstract}
Feedback-driven quantum reservoir computing has so far been studied primarily in gate-based architectures, motivating alternative scalable, hardware-friendly physical platforms. Here we investigate a linear-optical quantum reservoir architecture for time-series processing based on multiphoton interference in a reconfigurable interferometer network equipped with threshold detectors and measurement-conditioned feedback. The reservoir state is constructed from coarse-grained coincidence features, and the feedback updates only a structured, budgeted subset of programmable phases, enabling recurrence without training internal weights. By sweeping the feedback strength, we identify three dynamical regimes and find that memory performance peaks near the stability boundary. We quantify temporal processing via linear memory capacity and validate nonlinear forecasting on benchmarks, namely Mackey–Glass series, NARMA$-n$ and non-integrable Ising dynamics. The proposed architecture is compatible with current photonic technology and lowers the experimental barrier to feedback-driven QRC for time-series analysis with competitive accuracy.

\end{abstract}
\section*{Introduction}
Reservoir computing (RC) is a neuro-inspired machine learning (ML) framework that exploits the high-dimensional state space of nonlinear dynamical systems while training only a lightweight readout, which makes the learning quick and stable \cite{cucchi2022hands,gauthier2021next, jaeger2001echo}. For a reservoir to perform temporal computation, it should exhibit two key properties, namely input separability and fading memory. Input separability is typically obtained by a nonlinear mapping of low dimensional inputs to a high dimensional feature space. Fading memory is the ability to retain information about past inputs while ensuring that their influence on the internal state decays smoothly over time, preventing sensitive dependence on initial conditions. Reservoirs with these properties can be implemented as artificial recurrent neural networks or as physical dynamical systems.

Within the class of physical dynamical systems, quantum systems draw attention due to their inherent nonlinearity and high dimensionality, giving rise to the framework of quantum reservoir computing (QRC) \cite{ fujii2017harnessing, ghosh2019quantum}. Because QRC uses the natural evolution of quantum systems for information processing, it is well suited to noisy intermediate-scale quantum (NISQ) devices, in contrast to many quantum ML approaches that depend on fault-tolerant computation \cite{wang2024comprehensive}. A key obstacle for temporal processing in QRC is measurement back-action. However, these measurements unavoidably disturb or destroy the reservoir state and lead to significant information loss. A promising route to mitigate this problem is feedback-based control \cite{kobayashi2024feedback}, which is similar in spirit to variational quantum algorithms \cite{cerezo2021variational} but avoids variational optimization. In such schemes, classical feedback that is a function of recent measurement results is used to steer programmable elements of the device and hence sustain tunable memory. While such schemes have been proposed for gate-based architectures \cite{kobayashi2024feedback, monomi2025feedback, wudarski2023hybrid}, their practical realization can be demanding, prompting interest in alternative physical platforms. One promising alternative platform where this approach can be implemented is found in photonics models of quantum interference in reconfigurable linear optical platforms, where photons coherently evolve through networks of beamsplitters and phase shifters. Such models have been of central importance, fuelling fundamental advances in photonic quantum technologies and being proposed for fault-tolerant photonic quantum computing \cite{knill2001scheme, bartolucci2023fusion} as well as a pathway to quantum advantage for specialized algorithms \cite{aaronson2011computational}. Beyond these foundational directions, the same framework has been mapped to application specific tasks such as molecular dynamics \cite{sparrow2018simulating, huh2015boson}, and graph problems \cite{deng2023solving}, and has also been used to enhance classical algorithms \cite{arrazola2018using}. Recently, ML applications have emerged in which the model acts as a static physical feature generator, for example in image classification tasks \cite{cimini2025large, gong2025enhanced, sakurai2025quantum, nerenberg2025photon}, suggesting its suitability as a substrate for reservoir-style temporal processing through a feedback control. 

In this work, we investigate a reconfigurable linear-optical QRC platform that is explicitly designed to (i) use experimentally accessible measurements and (ii) offer a controllable fading-memory profile via structured, budgeted feedback actuation. The reservoir consists of an $M$-mode interferometer mesh driven by a scalar input sequence, implemented as a Galton-board--like wedge of Mach--Zehnder interferometers (MZIs) embedded in a larger static random mixing network. At each discrete time step $k$, the current sample is encoded by modulating only a shallow input block. The photonic state then propagates through a mesh whose programmable subset has already been updated from the most recent measurement cycle, rendering the evolution explicitly history dependent. Crucially, we take the reservoir state to be a vector of coarse-grained coincidence features obtained with threshold (on--off) detectors, which removes the need for photon-number-resolving detection while retaining a high-dimensional nonlinear feature map. A fixed random linear map transforms the most recent coincidence vector into phase updates applied only to the MZIs inside the Galton wedge, while all remaining MZIs remain static. This partial reconfiguration implements a physically realistic feedback mechanism that injects recurrence without requiring training of internal weights and without demanding full-mesh reprogramming at every step. We characterize both dynamics and computation by systematically sweeping the feedback gain. This reveals three qualitative regimes:an input-responsive stable phase, and an unstable phase and a feedback-dominated phase. Consistent with the edge-of-chaos hypothesis, memory performance peaks near the stability boundary \cite{langton1990computation, nishioka2022edge}. We then quantify temporal processing via linear memory capacity and validate forecasting performance on Mackey--Glass (MG), NARMA$-n$, and non-integrable 1-D Ising chain dynamics benchmarks. Finally, we discuss experimentally relevant effects, including loss and finite measurement sampling, to connect the idealized model with near-term photonic implementations.

\section*{Photonic Platform}

We briefly review general case of linear optical quantum processor, composed of single photons as inputs, a linear optical network, and detectors. Let us consider \(N\) indistinguishable, non-interacting photons evolving through an \(M\)-mode linear optical network described by a unitary \(\mathcal V\in U(M)\). Let \(\ket{S}=\ket{s_1\,s_2\,\dots\,s_M}\) be the input Fock state with \(\sum_{j=1}^M s_j=N\). The output Fock basis is enumerated as \(\{\ket{Q_\ell}\}_{\ell=1}^{\text{d}_{\text{pnr}}}\), where each \(\ket{Q_\ell} = \ket{q^{(\ell)}_1 q^{(\ell)}_2 \dots q^{(\ell)}_M}\) satisfying \(\sum_{j=1}^M q^{(\ell)}_j=N\) and \(\text{d}_{\text{pnr}}=\binom{M+N-1}{N}\). The transition probability is 
\begin{equation}
p^{(\mathcal V)}\!\left(\ket{S}\!\to\!\ket{Q_\ell}\right)
= \frac{\bigl|\operatorname{per}\!\bigl(\mathcal{V}_{Q_\ell,S}\bigr)\bigr|^2}
{\prod_{i=1}^{M} s_i!\;\prod_{j=1}^{M} q^{(\ell)}_j!}\,,
\label{eq:prob}
\end{equation}
where $\operatorname{per}(A)=\sum_{\sigma\in S_N}\prod_{i=1}^{N} A_{i,\sigma(i)}$ is the matrix permanent (it appears because amplitudes over all input–output assignments add with the same sign for bosons). The submatrix \(\mathcal{V}_{Q_\ell,S}\) is obtained by repeating the \(i\)-th column of \(\mathcal V\) exactly \(s_i\) times and the \(j\)-th row exactly \(q^{(\ell)}_j\) times. In practice, resolving full output distributions typically requires photon-number-resolving (PNR) detectors due to collision, i.e. two or more photons in the same mode. However, threshold detectors that only distinguish ``vacuum'' from ``one or more photons'' may provide informative statistics even in the presence of collisions.  In this case, the measurement outcome is a binary click pattern $\vec {r} = (r_1,\dots,r_M)\in\{0,1\}^M$, where $r_j=1$ denotes a click in mode $j$ and $r_j=0$ denotes no click. The probability of observing a particular threshold pattern  $\vec {r}$  can be obtained by coarse-graining the PNR distribution over all Fock states. Explicitly,
\begin{equation}
	p^{(\mathcal V)}(\vec r \mid \ket{S})
	=
	\sum_{\ell=1}^{\text{d}_{\text{pnr}}}
	p^{(\mathcal V)}\!\left(\ket{S}\!\to\!\ket{Q_\ell}\right)
	\prod_{j=1}^M
	\Bigl[\mathbf 1\bigl(q_j^{(\ell)} > 0\bigr)\Bigr]^{r_j}
	\Bigl[\mathbf 1\bigl(q_j^{(\ell)} = 0\bigr)\Bigr]^{1-r_j},
	\label{eq:threshold-from-fock}
\end{equation}
where $\mathbf 1[\cdot]$ denotes the indicator function, equal to $1$ if its argument is true and $0$ otherwise. The same quantity can also be expressed in closed form in terms of the Bristolian function \cite{bulmer2022threshold}.  The number of distinct threshold binary pattern is $\text{d}_{\text{thr}} = \sum_{w=1}^{N} \binom{M}{w}$ corresponding to all binary strings $\vec {r}$ with Hamming weight $w$. 

This threshold features can further be compressed into coincidence features for the sake of experimental simplicity, while still capturing a large amount of structure. Using multichannel time tagging, at each discrete time step $k$ one can obtain a vector of cross-mode coincidences in lexicographic order, $\mathcal{C}_k = \bigl(C_{ij,k}\bigr)_{1 \le i < j \le M} \in \mathbb{R}^{\text{d}^{\text{c}}_{\text{thr}}},$ with $ \text{d}^{\text{c}}_{\text{thr}} = \binom{M}{2}$ (number of pairwise coincidence features), where $C_{ij,k}$ is the coincidence between channels $i$ and $j$, and it is given by
\begin{equation}
	C_{ij,k}
	\;=\;
	\sum_{\vec r \in \{0,1\}^M}
	r_i\,r_j\;
	p_k^{(\mathcal V)}(\vec r \mid \ket{S}),
	\qquad i<j,
	\label{eq:Cij-from-pr}
\end{equation}
where $p^{(\mathcal V)}_k(\vec r \mid \ket{S})$ corresponds to threshold distribution at the time step $k$. 

Hence, each $C_{ij,k}$ is obtained by coarse-graining the number-resolved statistics over all Fock states that simply produce simultaneous clicks in modes $i$ and $j$. When data are encoded into the interferometer (for example, by modulating beamsplitter angles), a linear optical reservoir with detectors projects nonlinearly input data onto the high-dimensional feature space. With a linear readout on these output features, the system can be used as an extreme learning machine, i.e., a memoryless reservoir.  In what follows, we use threshold detection, which removes the necessity of using PNR detection, together with coincidence-based classical features for both readout and feedback.

\subsection*{Model Architecture}

\begin{figure}[!htbp]
\centering
\includegraphics[width=0.85\linewidth]{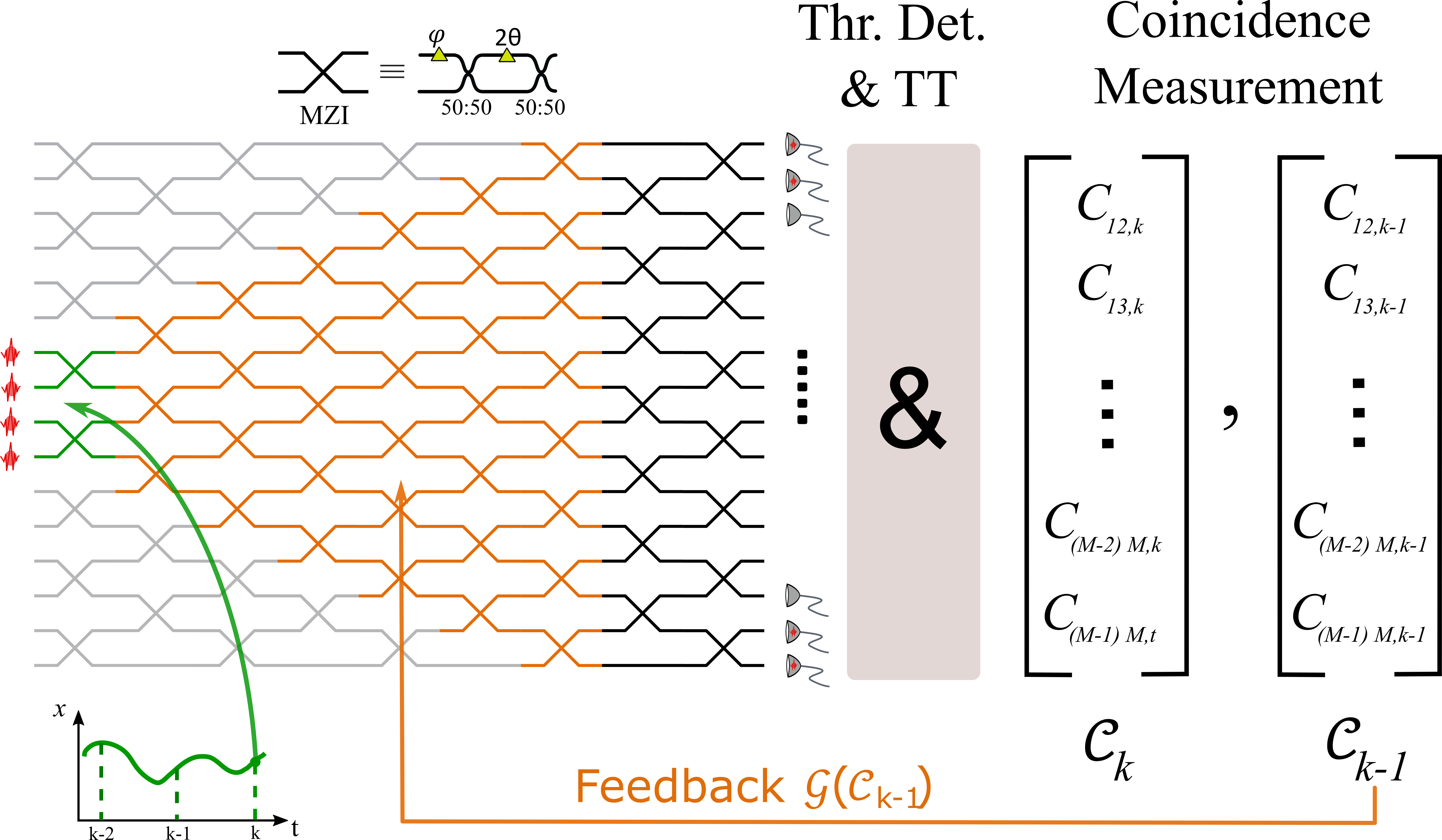}
\caption{Feedback-driven linear optical reservoir for temporal processing. Schematic at cycle $k$: the input sample $x_k$ is encoded in a small input block of MZIs (green), cross-mode coincidence statistics from the most recent coincidence measurement in lexicographic order ($C_{12,k-1}, C_{13,k-1}, \dots, C_{(M-1) M ,k-1}$) program the Galton wedge of MZIs (orange) through the feedback. The downstream static layers (black) provide fixed output mixing and can be considered as the reservoir part of the architecture. MZIs in the upper-left and lower-left corners that remain unilluminated for this input configuration are shown in gray. Crossings denote MZIs with yellow triangles representing phase shifters.}
\label{fig:schematic}
\end{figure}

In this section, we introduce a representative linear-optical QRC architecture, modeled as an $N$-input $M$-output-mode reconfigurable mesh that is driven by a scalar input sequence \{$x_k$\} and a classical feedback loop, see Fig.~\ref{fig:schematic}. The mesh contains a series of reconfigurable MZIs of the form
\begin{equation}
	R(\theta,\varphi)
	=
	\begin{bmatrix}
		e^{\mathrm{i}\varphi}\cos\theta & -\sin\theta \\
		e^{\mathrm{i}\varphi}\sin\theta & \phantom{-}\cos\theta
	\end{bmatrix},
\end{equation}
controlled by two parameters $(\theta,\varphi)$ per MZI. We divide the mesh into three parts: two ``input'' MZIs in the first layer that mix a central four-mode block, a central Galton-style fan-out wedge of ``feedback'' MZIs, and the post Galton MZIs, which are held at fixed random settings and provide static mixing, i.e. the reservoir part of the architecture. For our chosen input block, the upper-left and lower-left corners of the mesh are never illuminated and could in principle be removed (or replaced by straight waveguides) without affecting the dynamics, see Fig. \ref{fig:schematic}. The interferometer architecture is a so-called fully connected rectangular interferometer \cite{sund2024hardware}, all input ports are connected to all output ports. This interferometer can be realized by reducing the number of input modes to the number of input photons $N$, while maintaining $M$ output modes with full connectivity. In practice, a rectangular interferometer can be embedded inside a Clements interferometer \cite{clements2016optimal}, with unused MZIs. While we focus on Galton-style layout with a central four-mode input block, the same scheme can be implemented on other reconfigurable linear optical meshes (e.g., different mode numbers, depths, or connectivity patterns).

The input encoder is shallow and depends only on the current sample $x_k$ through an input strength $\alpha_{\mathrm{in}}$. We use two input MZIs in the first layer that mix the central block of modes, and modulate their mixing angles in a push--pull fashion around the balanced point $\theta = \pi/4$. Denoting the two input MZIs internal phases by $\theta^{(\mathcal{I})}_{1,k}$ and $\theta^{(\mathcal{I})}_{2,k}$ at a discrete time $k$, we set
\begin{equation}
	\theta^{(\mathcal{I})}_{1,k} = \frac{\pi}{4} + \alpha_{\mathrm{in}} x_k,
	\qquad
	\theta^{(\mathcal{I})}_{2,k} = \frac{\pi}{4} - \alpha_{\mathrm{in}} x_k,
\end{equation}
while their external phases are kept fixed.

The feedback block depends on a recent-history vector via a linear map $\mathcal{G}$ and a feedback strength $\alpha_{\mathrm{fb}}$, and reprograms only the MZIs inside the Galton wedge, leaving all remaining MZIs at their static random settings. We map the most recent cross-mode coincidence vector \(\mathcal{C}_{k-1}\) through a linear map \(\mathcal{G} : \mathbb{R}^{\text{d}^{\text{c}}_{\text{thr}}} \to \mathbb{R}^{2R_{\mathrm{fb}}}\), where \(R_{\mathrm{fb}}\) is the number of MZIs in the Galton wedge. We first multiply \(\mathcal{C}_{k-1}\) by a fixed random matrix \(\mathrm{V}_{\mathrm{fb}} \in \mathbb{R}^{2R_{\mathrm{fb}} \times \text{d}^{\text{c}}_{\text{thr}}}\) with entries \(\mathrm{V}_{ij} \stackrel{\text{i.i.d.}}{\sim} \mathcal{N}\!\bigl(0, 1/d\bigr)\) to obtain \(\mathbf{h}_k = \mathrm{V}_{\mathrm{fb}} \,\mathcal{C}_{k-1}\). We then convert \(\mathbf{h}_k\) into mesh-control amplitudes via \(\mathbf{a}_k = \tfrac{\pi}{4}\mathbb{1} + \alpha_{\mathrm{fb}}\,\mathbf{h}_k\), split this as \(\mathbf{a}_k = \bigl[\mathbf{a}_k^{(\theta)},\,\mathbf{a}_k^{(\varphi)}\bigr]^{\top}\), and map to physical parameters \(\boldsymbol{\theta}_k = \mathbf{a}_k^{(\theta)}\) and \(\boldsymbol{\varphi}_k = 4\,\mathbf{a}_k^{(\varphi)}\), with \(\alpha_{\mathrm{fb}}\) setting the feedback strength. These are assigned to the Galton wedge, while the remaining ones stay at their static random settings, so \(\mathcal{G}(\mathcal{C}_{k-1})\) specifies only the programmable subset. While we use a fixed untrained $\mathrm{V}$ in this work, the architecture naturally supports budgeted sparse actuation in which only a subset of Mach–Zehnder interferometers within the Galton wedge is dynamically controlled and the remaining elements are kept static.

In the training phase, for a lenght-$T$ run, the output layer is trained using Ridge regression with a regularization parameter of $\beta=10^{-11}$ using the standardized features $\mathcal{C}'_k$. At test phase, we reuse the same training statistics and produce predictions $\hat {y}_k = \mathbf{W}^\top\mathcal{C}'_k + b$, where $\mathbf{W}$ represents the output weights and $b$ is the bias term. The perfomance is evaluated through the normalized mean-squared error (NMSE), expressed as follows

\begin{equation}
	\mathrm{NMSE}
	= \dfrac{1}{L_e}\frac{\sum_{k} \bigl(\hat{y}_k-y_k\bigr)^2}
	{\sigma_y^{2}},
	\label{NMSE}        
\end{equation}
where  $\sigma_y$ is the standard deviation of y$_k$ over the evaluation window $L_e$. Lower NMSE indicate closer agreement between $\hat{y}_k$ and the ground-truth target y$_k$.

In summary, at each discrete time step $k>0$, the input sample $x_k$ is first encoded by the shallow block acting on the fixed input state $\ket{S}$ by modulating only the two input MZIs around the balanced point. The resulting state then propagates through the feedback-programmed Galton wedge, whose configuration depends on the previous coincidence vector $\mathcal{C}_{k-1}$. Finally, the state is mixed by the static block, which completes the interferometer, determines the output distribution and thereby produces a new coincidence vector $\mathcal{C}_k$. A single time step realizes the effective state update that depends on the current input $x_k$ and on the past, through the history implicit in the coincidence vectors. The interplay of multiphoton interference, coarse-grained coincidence measurements, and measurement-conditioned reprogramming of a Galton wedge thus induces a high-dimensional, nonlinear state update with a tunable fading-memory profile, precisely the ingredients that make physical reservoirs effective. Although we study this mechanism in a specific Galton-style mesh, the same design principles apply broadly to reconfigurable linear optical interferometers with different sizes, connectivities, and detection layouts, providing a general template for linear optical quantum reservoir computing for temporal processing.
  
\subsection*{Experimental feasibility of the design}
We now assess the viability of implementing our proposal with current photonic technology. We first focus on an experimentally realistic scalable scheme. Although the model itself does not impose a formal scaling limit in either $M$ or $N$, practical challenges emerge as systems grow, since interconnect complexity increases, optical loss accumulates, and phase stability becomes harder to maintain across the entire mesh. Integrated photonics helps mitigate these issues by offering low propagation loss, intrinsic phase stability and dense, programmable control over MZIs, for example through thermo-optic or electro-optic phase shifters implemented directly on chip  \cite{maring2024versatile, arrazola2021quantum}. It is noteworthy to point out that, in principle, the framework could equivalently be implemented with any suitable linear optical reservoirs.

We next consider two aspects that most strongly affect experimental performance: (i) photon loss and photon distinguishability, and (ii) the fidelity of input encoding and feedback control, since the mesh relies on accurate and reconfigurable phase shifters.

\textit{Photon loss and distinguishability:} 
Assuming uniform loss for all input–output configuration, we absorb it into a single efficiency parameter \cite{oszmaniec2018classical}. With per mode efficiency \(\eta_{\mathrm{eff}}\), the threshold click probability conditioned on an output Fock state $\ket{Q_\ell}$ is $ \Pr\!\big(c_i=1 \mid \ket{Q_\ell}\big) = 1 -\,\bigl(1-\eta_{\mathrm{eff}}\bigr)^{\,q_i^{(\ell)}}
\label{phloss}$. This expression replaces the indicator function in Eq. \ref{eq:threshold-from-fock} to incorporate losses, explicitly $p^{(\mathcal V, \eta_{\mathrm{eff}})}(\vec r \mid \ket{S}) = \sum_{\ell=1}^{d_{\mathrm{pnr}}} p^{(\mathcal V)}\!\bigl(\ket{S}\!\to\!\ket{Q_\ell}\bigr)\,
\prod_{j=1}^M \bigl[\Pr(c_j = 1 \mid \ket{Q_\ell})\bigr]^{r_j} \bigl[1 - \Pr(c_j = 1 \mid \ket{Q_\ell})\bigr]^{1-r_j}.$ In the few-photon regime we consider, uniform loss acts to a good approximation as an overall linear reduction of the coincidence feature, so its effect can largely be absorbed into an appropriate choice of the feedback gain $\alpha_{\mathrm{fb}}$. In addition, we also assume indistinguishable photons, consistent with reported indistinguishability at the level of about \(99\%\) \cite{paesani2020near,somaschi2016near}. However, when partial distinguishability is not negligible, the more general models such as multipermanent model \cite{tichy2015sampling} can be used to account for it.

\textit{Input encoding and feedback control:}
The primitive elements of the mesh, MZIs, can be tuned close to ideal behavior with \(>\!60\,\mathrm{dB}\) extinction \cite{wilkes201660}. At the level of full meshes, 100 Haar unitaries have been implemented  with an average fidelity of \(0.999\pm0.001\) \cite{carolan2015universal}, while 10000 arbitrary linear transformations on a silicon photonic have recently been realized with record-high fidelity of \(0.99997\pm0.0002\) limited mainly by the measurement equipment \cite{moralis2024perfect}. These results support our assumption that an input encoding and a feedback mesh can be realized on near term reprogrammable photonic platforms with near-unity accuracy.

\section*{Performance Evaluation}
We quantify linear memory capacity of the system and evaluate the predictive capability of our photonic QRC on MG series benchmarks, NARMA-$n$ and  under the assumption of infinite measurements $N_{\text{m}} \rightarrow \infty$. Under this assumption, statistical uncertainty in measurements vanishes, and performance is governed solely by the system’s intrinsic dynamics. All the simulation of the system have been carried out using the Python programming language. Each dataset comprises 2200 time steps with a 200-step washout, followed by 1000 training steps and 1000 test steps. We choose system size of $(M,N)= (16,4)$ to have a balance between expressivity and current experimental capabilities.

\subsection*{Memory Capacity}

We quantify the reservoir’s temporal processing using the linear memory capacity (MC), which measures how well a linear readout can reconstruct past inputs under i.i.d.\ driving. Unless stated otherwise, we evaluate delays $1\le \tau \le 25$ and average results over 30 independent realizations (reporting mean values only).

To estimate linear memory, we draw a scalar drive $\{s_k\}_{k=1}^{T_{\mathrm{tot}}}$ i.i.d.\ from $[0,1]$. For each delay $\tau\in\{1,\dots,25\}$, we define the linear target $y_{k,\tau}=s_{k-\tau}.
$ A linear readout is trained to produce predictions $\widehat{y}_{k,\tau}$. The linear memory capacity at delay $\tau$ is computed as
\begin{equation}
	\mathrm{MC}(\tau)=1-\mathrm{NMSE}\!\left(\widehat{y}_{:,\tau},\,y_{:,\tau}\right).
\end{equation}
Finally, we define the total linear memory capacity as
\begin{equation}
	\mathrm{MC}^{\mathrm{tot}}=\sum_{\tau=1}^{\mathcal T}\mathrm{MC}(\tau)
\end{equation}

First, we investigate the effectiveness of feedback connections. With the input gain fixed at $\alpha_{\mathrm{in}}=0.001$, we examine how linear memory, $\mathrm{MC}_1(\tau)$, depends on the feedback strength $\alpha_{\mathrm{fb}}$, see Fig. \ref{fig:MC} \textbf{a}. 
For \(\alpha_{\mathrm{fb}}=2.2\), \(\mathrm{MC}_1(\tau)\) remains close to unity up to \(\tau\approx 12\) before decaying, showing that input histories are effectively accumulated in the coincidence outcomes and carried forward by the feedback. Increasing the feedback to \(\alpha_{\mathrm{fb}}=2.4\) slightly suppresses short-delay memory, reflecting stronger recurrent influence relative to fresh input. For \(\alpha_{\mathrm{fb}}=3.2\), the memory curve becomes noticeably flatter and reduced, and for \(\alpha_{\mathrm{fb}}=4.6\) the capacity is almost zero at all delays, indicating that the coincidence patterns are largely insensitive to the input history. 

\begin{figure*}[!htbp]

	\centering
		\includegraphics[scale=0.37]{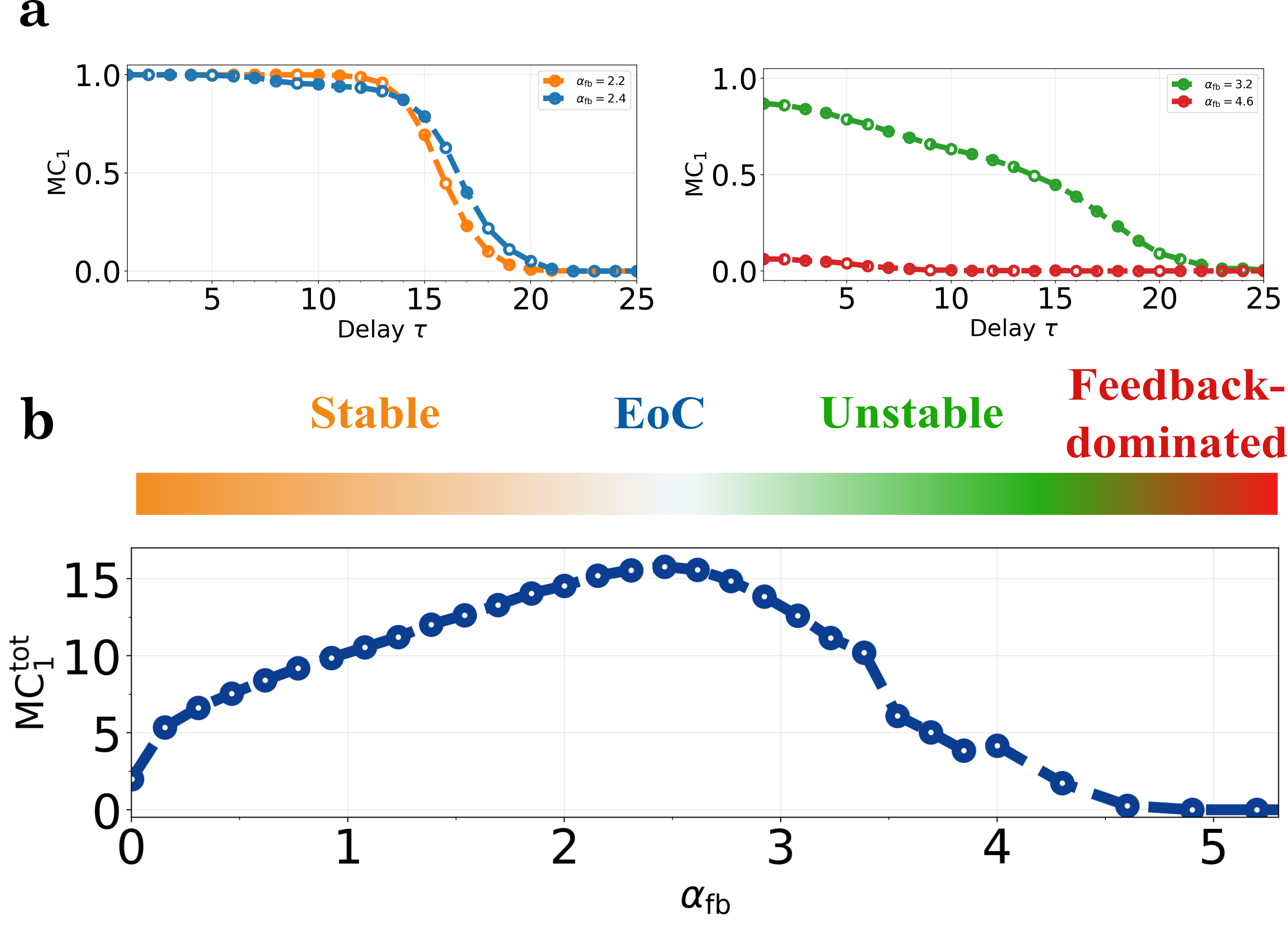}
	\caption{%
		Linear memory performance of the feedback-driven reconfigurable linear optics as a function of delay for different feedback strengths for a fixed device size $(M,N)=(16,4)$ and fixed input strength $\alpha_{\mathrm{in}}$. \textbf{a} $\alpha_{\mathrm{fb}} = 1.5$ (blue), $\alpha_{\mathrm{fb}}=1.8$ (orange),  $\alpha_{\mathrm{fb}} = 2.25$ (green) and $\alpha_{\mathrm{fb}}=2.75$ (red) \textbf{b} Total linear capacity, MC$^{^{\text{tot}}}_1$, as a function of feedback strength $\alpha_{\mathrm{fb}}$.}
			\label{fig:MC}
\end{figure*}

In Fig.~\ref{fig:MC}\,\textbf{b}, we plot the total linear memory capacity $\mathrm{MC}_1^{\text{tot}}$ versus $\alpha_{\mathrm{fb}}$, which reveals three qualitatively distinct regimes. For weak to moderate feedback ($0 < \alpha_{\mathrm{fb}} \lesssim 2.4$), the capacity generally increases with $\alpha_{\mathrm{fb}}$, where the reservoir gains useful recurrence while remaining well behaved and input-responsive. For intermediate feedback $2.4 \lesssim \alpha_{\mathrm{fb}} \lesssim 4.7$, the capacity decreases as the feedback term increasingly dominates the influence of the input. Beyond this range ($\alpha_{\mathrm{fb}} \gtrsim 4.7$) the system resides in a strongly feedback-dominated phase in which the coincidences quickly lose any linearly decodable dependence on past inputs and $\mathrm{MC}_1^{\text{tot}}$ is nearly zero. The optimal operating point lies just below the onset of this unstable behaviour, in close proximity to the boundary between the input-responsive and unstable phases, corresponds to the EoC for this quantum reservoir.

\subsection*{Predictive Performance}
Having  demonstrated the linear memory performance facilitated by the feedback connections, we now turn to the predictive performance of our QRC protocol on three standard benchmarks.

\paragraph{Mackey--Glass.}
We first consider the Mackey--Glass (MG) delay-differential equation \cite{mackey1977oscillation}, widely used in reservoir computing:
\begin{equation}
	\dot{Q}(t) \;=\; \alpha\,\frac{Q(t-\tau)}{1 + Q(t-\tau)^{\beta}} \;-\; \gamma\,Q(t),
\end{equation}
with $(\alpha,\beta,\gamma)=(0.2,10,0.1)$ and $\tau=17$, for which the dynamics exhibits a chaotic attractor (for $\tau \gtrsim 16.8$).
We discretize time with step $\Delta t=1$, so that one MG update $Q(t)\!\to\!Q(t+\Delta t)$ corresponds to one reservoir cycle $k\!\to\!k+1$.
The resulting scalar stream is obtained by normalizing the samples of $Q(t)$ to $x_k \in [0,1]$ and encoding $x_k$ into the processor.
For a prediction horizon $\tau_f$, the readout is trained to reproduce the target $y_k = x_{k+\tau_f}$.

\begin{figure*}[!htbp]
	\centering
	\includegraphics[scale=0.6]{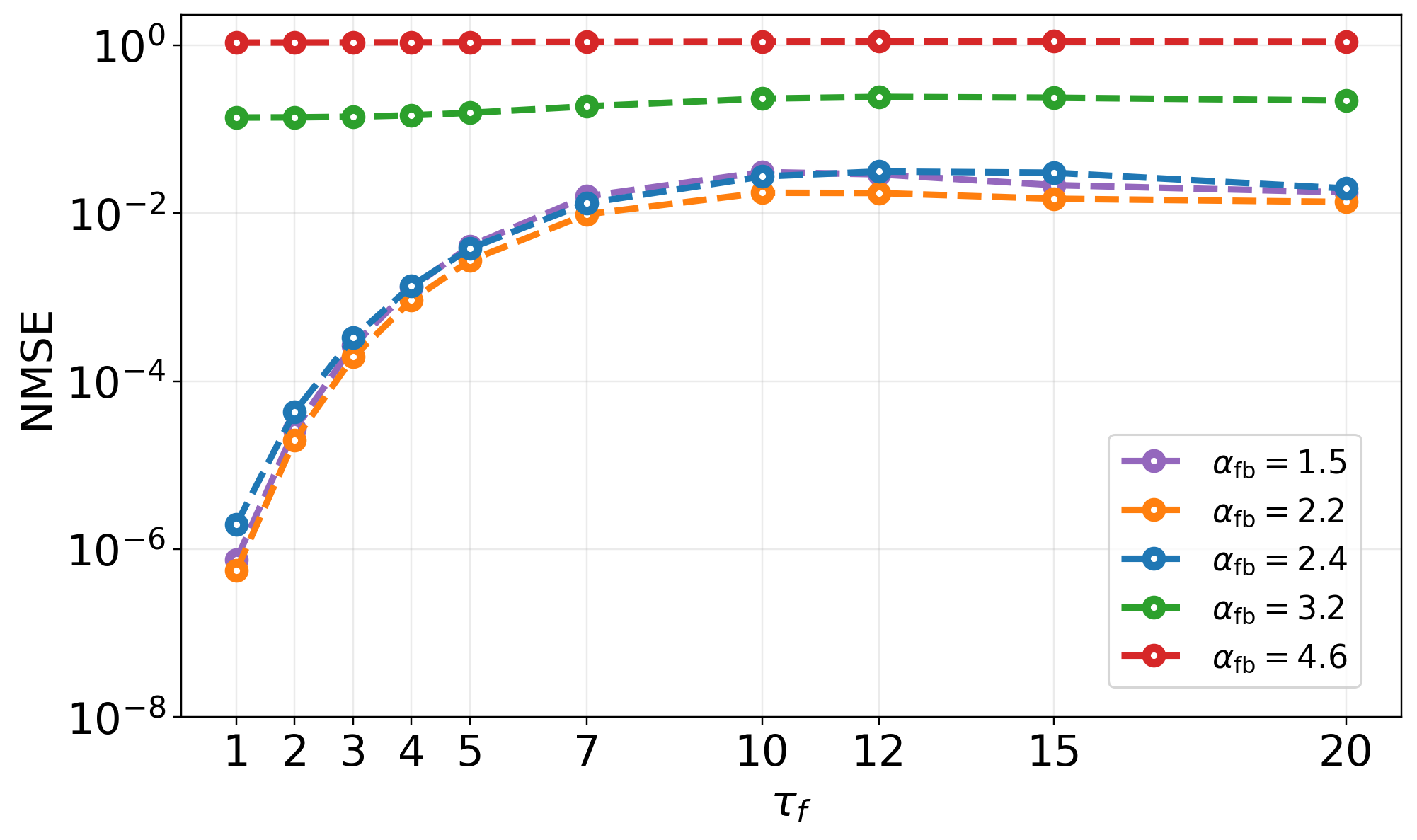}
	\caption{Mackey--Glass prediction performance. Mean normalized mean-square error (NMSE, log scale) as a function of the prediction horizon $\tau_f$. Each curve corresponds to a different feedback gain $\alpha_{\mathrm{fb}}$ and reports the mean over 30 different realizations.}
	\label{fig:mg}
\end{figure*}
\begin{figure*}[!htbp]
	\centering
	\includegraphics[scale=0.6]{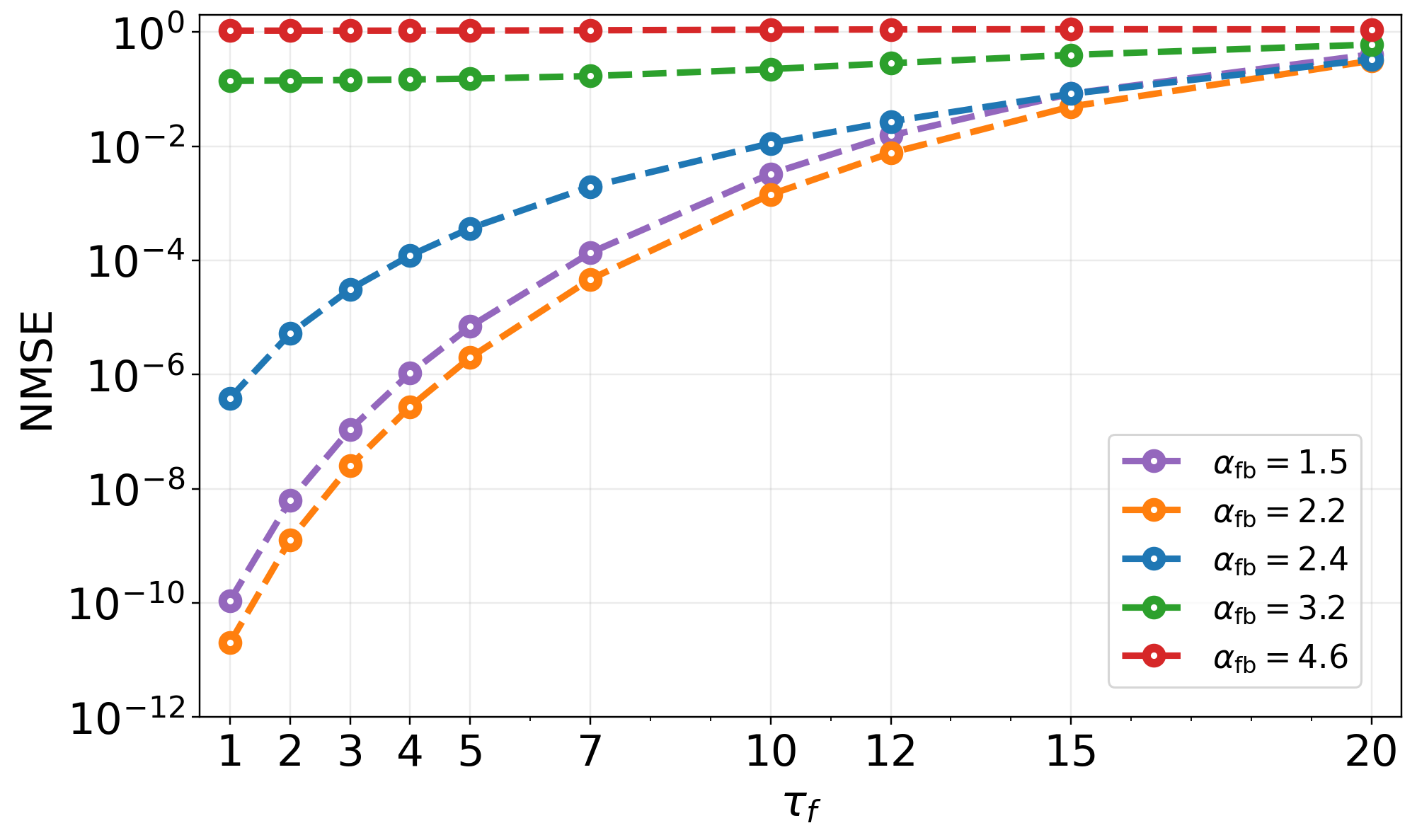}
	\caption{Non-integrable quantum Ising forecasting. Mean NMSE (log scale) versus prediction horizon $\tau_f$ for the central-spin signal $\langle \sigma_3^{z}(t)\rangle$ (mapped to $x_k\in[0,1]$). Each curve corresponds to a feedback gain $\alpha_{\mathrm{fb}}$ and shows the mean over 30 realizations.}
	\label{fig:ising}
\end{figure*}

Figure~\ref{fig:ising} shows that the prediction error increases with the horizon $\tau_f$, reflecting the progressive loss of predictability in the interacting quantum dynamics. The feedback gain $\alpha_{\mathrm{fb}}$ again plays a central role: intermediate values $\alpha_{\mathrm{fb}}\approx 2.2$--$2.4$ minimize NMSE across horizons, while overly strong feedback degrades performance and drives the readout toward a high-error regime.

Figure~\ref{fig:mg} shows that the prediction error increases with the forecast horizon $\tau_f$, as expected for MG dynamics.
The feedback gain $\alpha_{\mathrm{fb}}$ strongly controls performance: intermediate gains around $\alpha_{\mathrm{fb}}\approx 2.2$--$2.4$ yield the lowest NMSE across horizons, whereas larger gains degrade performance and rapidly approach a large-error regime, consistent with feedback-driven instability.
Smaller gains remain stable but provide slightly reduced short-horizon accuracy compared with the optimal intermediate regime.

\paragraph{NARMA$-n$.}
As a second benchmark, we study the NARMA$-n$ task, a standard test of combined memory and nonlinearity in reservoir computing.
Given a scalar input stream $\{x_k\}$ and output $\{y_k\}$, the target obeys
\begin{equation}
	y_{k+1}
	= \alpha\, y_k
	+ \beta\, y_k \sum_{j=0}^{n-1} y_{k-j}
	+ \gamma\, x_k\, x_{k-n+1}
	+ \delta,
\end{equation}
with $(\alpha,\beta,\gamma,\delta)=(0.3,\,0.05,\,1.5,\,0.1)$.
The input samples $x_k$ are drawn i.i.d.\ from a uniform distribution on $[0,0.5]$.
We fix the input strength to $\alpha_{\mathrm{in}} = 0.001$ and sweep the feedback strength $\alpha_{\mathrm{fb}} \in \{1.5,\,2.2,\,2.4,\,3.2,\,4.6\}$.
For each $\alpha_{\mathrm{fb}}$, we run the protocol for NARMA-7 and NARMA-10 and average the test NMSE over 20 random instances.

\begin{figure*}[!htbp]
	\centering
	\includegraphics[scale=0.6]{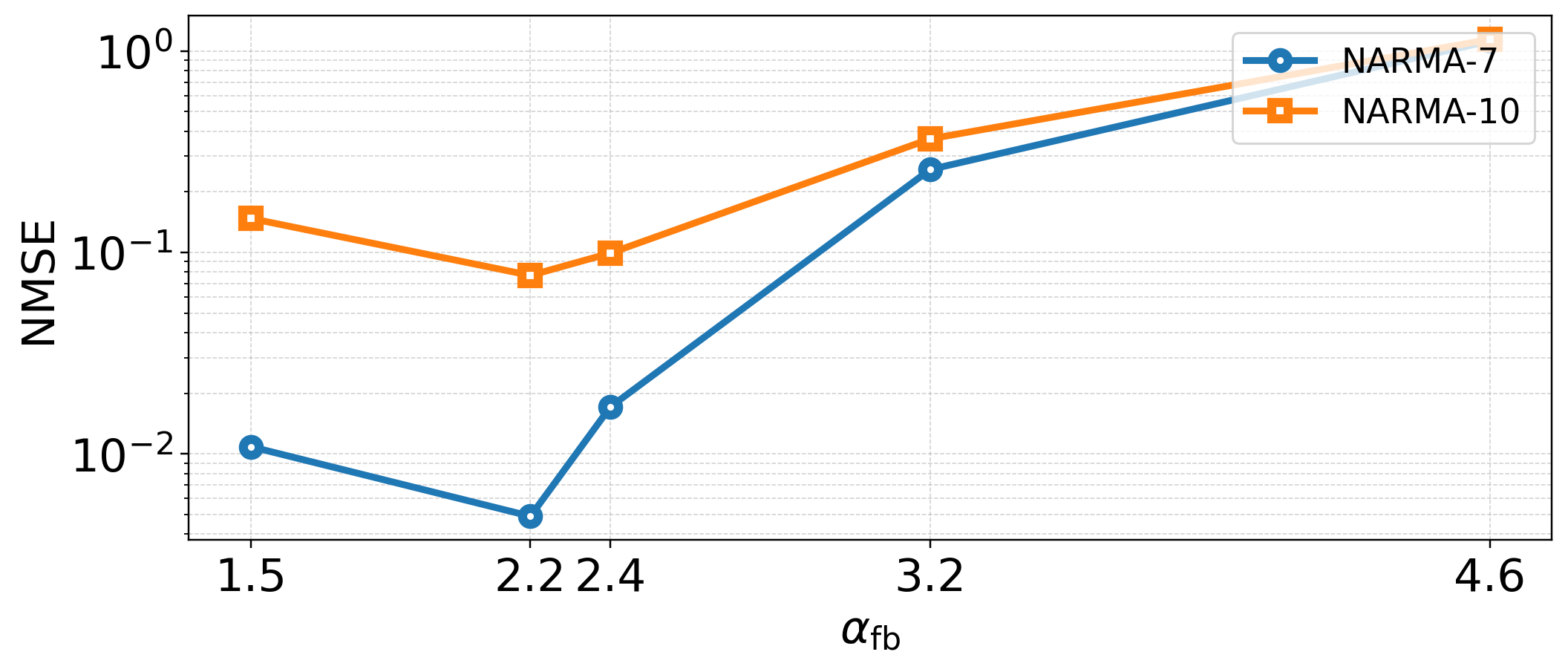}
	\caption{%
		NMSE for NARMA-7 (circles) and NARMA-10 (squares) as a function of feedback strength $\alpha_{\mathrm{fb}}$. The input strength is fixed to $\alpha_{\mathrm{in}} = 0.001$.}
	\label{fig:narma}
\end{figure*}

The results are summarized in Fig.~\ref{fig:narma}.
For NARMA-7, the best performance is obtained around $\alpha_{\mathrm{fb}} \approx 2.2$, where the NMSE reaches $\mathrm{NMSE}_7 \approx 4.9\times10^{-3}$, while both weak feedback ($\alpha_{\mathrm{fb}}=1.5$) and strongly overdriven feedback ($\alpha_{\mathrm{fb}} \geq 4.6$) degrade the predictive accuracy.
For the more demanding NARMA-10 task, the error reaches $\mathrm{NMSE}_{10} \approx 7.6\times 10^{-2}$ in the optimal regime, again near $\alpha_{\mathrm{fb}} \approx 2.2$.
At large feedback strengths the NMSE increases by almost an order of magnitude, indicating that the reservoir is pushed into a regime that is no longer well matched to the statistics of the NARMA input.

\paragraph{1-D quantum Ising chain forecasting.}
Finally, we forecast a quantum time series generated by a one-dimensional Ising chain governed by
\begin{equation}
	\mathcal{H} = -J \sum_i \sigma_i^{z}\sigma_{i+1}^{z} + h_x \sum_i \sigma_i^{x} + h_z \sum_i \sigma_i^{z}.
\end{equation}
Here $\sigma_i^{x}$ and $\sigma_i^{z}$ are the Pauli $x$ and $z$ matrices at site $i$, $h_x$ and $h_z$ denote the transverse and longitudinal fields, and $J$ is the nearest-neighbour coupling.
We simulate a chain of size $N=5$ with $J=1$, $h_x=1.05$, and $h_z=-0.5$ (a non-integrable setting with spectral statistics consistent with quantum chaos \cite{banuls2011strong}). From the resulting dynamics we record the central-spin observable $m(t)=\langle \sigma_3^{z}(t) \rangle$, sampled at times $t_k=k\,\Delta t$ with $\Delta t = 0.05$, which serves as the target time series for forecasting. One update $t_k\!\to\!t_{k+\Delta t}$ corresponds to one reservoir cycle $k\!\to\!k+1$.
The sampled signal $m_k$ is mapped to $x_k\in[0,1]$ before being encoded into the processor.

Figure~\ref{fig:ising} shows a clear transition from accurate to ineffective forecasting as the horizon increases.
Near EoC feedback stregnth ($\alpha_{\mathrm{fb}}\approx 2.2$) achieves the lowest errors and remains stable across horizons, whereas stronger and  feedback pushes the reservoir into an overdriven regime:
at $\alpha_{\mathrm{fb}}=4.6$ the NMSE is close to unity for essentially all $\tau_f$, indicating that the readout performs little better than a variance-level baseline.

These benchmarks show that the reservoir attains competitive predictive performance on both classical (MG and NARMA-$n$) and quantum (Ising) time-series tasks.
Across tasks, the feedback strength $\alpha_{\mathrm{fb}}$ acts as an effective control knob that tunes the trade-off between stability and rich nonlinear dynamics, with an intermediate regime (near the EoC, $\alpha_{\mathrm{fb}} \approx 2.2$) yielding the best overall predictive accuracy.
 
\subsection*{Finite measurement ensemble}
In this section we consider a more realistic setting in which cross-mode coincidences are estimated from a finite number of measurement ensembles $N_m$. In practice, the expectation values are estimated by averaging over a large ensemble of measured values, which inherently introduces statistical uncertainty. In this system, these statistical fluctuations in measurements is crucial, since they do not only influence the calculation of the final output $\hat{y}_k$ but also perturb the dynamics of the system itself, as the statistical noise is concurrently fed back inside the \(\mathcal{C}_{k-1}\). The statistical uncertainty scales as $\mathcal{O}(1/\sqrt{N_m})$, constraining the accuracy of the feedback based quantum reservoirs. In addition to that, when $N_m$ is insufficiently large, statistical fluctuations dominate over the impacts of the input operation, leading to inadequate information encoding. However, this issue can be addressed either by increasing the number of measurements $N_m$ or by amplifying the input weight $\alpha_{in}$ preventing the dominance of statistical fluctuations.


\begin{figure*}[!htbp]
	
	\centering
	\includegraphics[scale=0.27]{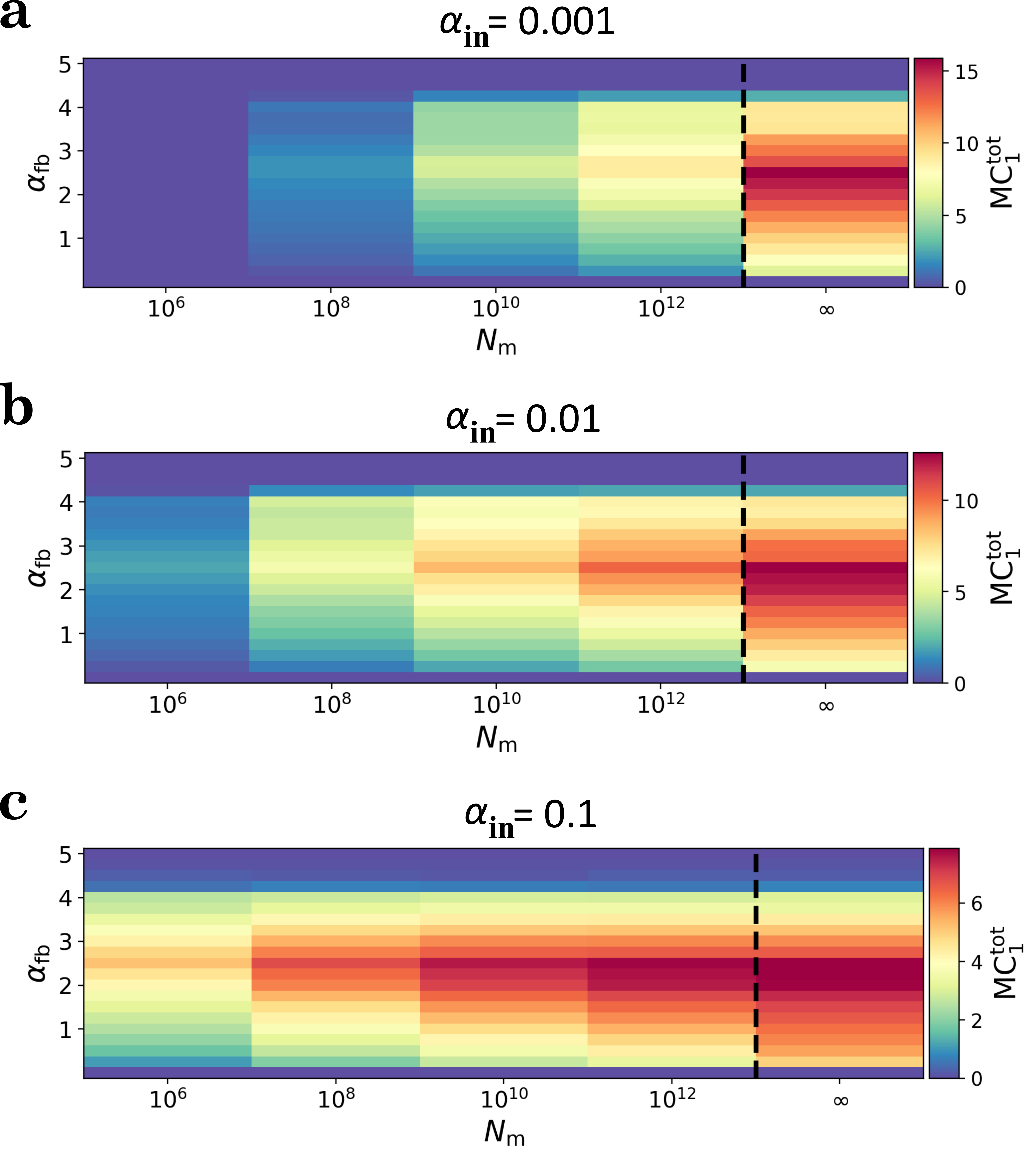}
	\caption{%
		Linear memory performance of the feedback-driven reconfigurable linear optics as a function of the number of measurements $N_\text{m}$ and feedback strength $\alpha_{\mathrm{fb}}$. The ideal capacities for $N_{\text{m}} \rightarrow \infty$ are also shown for reference. The input strength is set to  \textbf{a} $\alpha_{\mathrm{in}} = 0.001$  \textbf{b} $\alpha_{\mathrm{in}} = 0.01$ and \textbf{c} $\alpha_{\mathrm{in}} = 0.1$.}
	\label{fig:MC}
\end{figure*}

Figure \ref{fig:MC} plots the total capacity, $\mathrm{MC}^{\mathrm{tot}}$, as a function of the number of measurements $N_m$ and the feedback strength $\alpha_{\mathrm{fb}}$, for three different input weights $\alpha_{\mathrm{in}}$. As a reference, the noise-free performance (corresponding to $N_m \rightarrow \infty$) is also shown. A key observation is that the sensitivity to $N_m$ depends strongly on the input weight $\alpha_{\mathrm{in}}$. When $\alpha_{\mathrm{in}}$ is small, $\mathrm{MC}^{\mathrm{tot}}$ stays close to zero for essentially all $\alpha_{\mathrm{fb}}$ up to $N_m=10^{8}$. Only for substantially larger $N_m$ does $\mathrm{MC}^{\mathrm{tot}}$ gradually become nonzero, at which point the three feedback-strength regimes re-emerge. Increasing the input weight to $\alpha_{\mathrm{in}}=0.01$ shifts the onset of nonzero $\mathrm{MC}^{\mathrm{tot}}$ to smaller $N_m$. For an even larger input strength, $\alpha_{\mathrm{in}}=0.1$, the total capacity approaches the ideal (expectation-value) result rapidly, reaching near-convergence already around $N_m \approx 10^{10}$. Overall, while statistical fluctuations can perturb QRC dynamics, their effect can be reduced by appropriately tuning the input strength $\alpha_{\mathrm{in}}$.

\section*{Conclusion}
We analyzed a feedback-driven photonic quantum reservoir computing platform for time-series processing based on multiphoton interference in a reconfigurable interferometer, coarse-grained coincidence readout with threshold detectors, and measurement-conditioned phase updates. The key architectural feature is a structured, budgeted feedback loop that reprograms only a targeted subset of interferometer phases (the Galton wedge), thereby inducing recurrence and tunable fading memory without training internal weights or requiring full-mesh reconfiguration at every step. By sweeping the feedback gain, we identified three dynamical regimes—an input-responsive stable regime, an unstable transition regime, and a feedback-dominated regime in which the readout becomes largely insensitive to the input history. Across these regimes, linear memory capacity peaks near the stability boundary, consistent with the edge-of-chaos picture. Beyond linear memory, we validated temporal processing performance on standard nonlinear forecasting benchmarks (Mackey--Glass and NARMA$-n$) as well as forecasting of non-integrable quantum Ising dynamics, finding that the same intermediate-feedback region that maximizes memory also minimizes prediction error across horizons. We further examined finite measurement ensembles, where shot noise perturbs not only the readout but also the closed-loop dynamics through the feedback channel. Our results show that this degradation can be compensated in a hardware-relevant way by adjusting the input gain, enabling recovery of near-ideal performance at feasible measurement budgets. These findings position reconfigurable photonic interferometers with threshold detection and feedback actuation as a practical, scalable route to feedback-driven QRC for temporal learning.

\section*{Acknowledgments}
The author declares no competing financial interest. During the completion of this work, we became aware of related work \cite{di2025time}.
\appendix 
\bibliographystyle{unsrtnat}
\bibliography{sample}

\end{document}